\RequirePackage{lineno} 
\documentclass[a4paper,10pt,onecolumn,twoside,english]{scrartcl}

\usepackage{n3ltemplate}
\usepackage[square,numbers]{natbib}
\usepackage{float,calc,ifthen,amsmath}
\usepackage{color}
\usepackage{hyperref}      
\hypersetup{
    colorlinks=false,       
    linkcolor=blue,          
    citecolor=blue,        
    filecolor=blue,      
    urlcolor=blue          
}
\newcommand{\ord}{\mbox{d}}

\newcommand{\opL}{\mbox{L}}

\newcommand{\matI}{\mathbf{I}}

\newcommand{\matq}{\mathbf{q}}
\newcommand{\matqh}{\mathbf{\hat{q}}}

\newcommand{\xh}{\hat{x}}
\newcommand{\yh}{\hat{y}}

\newcommand{\qvec}{{\bf q}}
\newcommand{\qavec}{{\bf q}^{+}}

\newcommand{\rr}{\right}
\newcommand{\lf}{\left}

\newcommand{\ep}{\epsilon}

\newcommand{\dtqvec}{\delta\tilde\qvec}
\newcommand{\dqvec}{\delta\qvec}

\begin{document}
%

\title{Adjoint-based linear analysis in reduced-order thermo-acoustic models} 

\author{Luca Magri$^{\ast}$ \& Matthew P. Juniper}

\Contact{Department of Engineering, University of Cambridge \\
					Trumpington Street, CB2 1PZ, Cambridge, U.K. \\
					$^{\ast}$	Corresponding author:  \href{mailto:lm547@cam.ac.uk}{lm547@cam.ac.uk}\\	
					}

\maketitle\thispagestyle{plain}


\pagestyle{fancy} 

\fancyhf{} 

\fancyhead[L]{} 
\fancyhead[C]{\sf Luca Magri \& Matthew P. Juniper \\} 
\fancyhead[R]{} 

\renewcommand{\headrulewidth}{0.0pt}
\fancyfoot[C]{\sf \thepage}

%
\newcommand{\ccolor}{black}
\begin{abstract}
This paper presents the linear theory of adjoint equations as applied to thermo-acoustics.
The purpose is to describe the mathematical foundations of adjoint equations for linear sensitivity analysis of thermo-acoustic systems, recently developed by Magri and Juniper (J. Fluid Mech. (2013), vol. 719, pp. 183--202).
This method is applied pedagogically to a damped oscillator, for which analytical solutions are available, and then for an electrically heated Rijke tube with a mean-flow temperature discontinuity induced by the compact heat source.
Passive devices that most affect the growth rate / frequency of the electrical Rijke-tube system are presented, including a discussion about the effect of modelling the mean-flow temperature discontinuity.
\end{abstract}
\section{Introduction}
In a thermo-acoustic system, if the heat release is sufficiently in phase with the acoustic pressure waves, then oscillations can be enhanced, sometimes with negative consequences on the overall performance \cite{Rayleigh1878,Lieuwen2005,Culick2006,Lieuwen2012}. 
Many theoretical techniques have recently been introduced in thermo-acoustics, making this field an exciting hub of cross-disciplinary methods and theories. 
These include non-normality \cite{Balasubramanian2008a,Balasubramanian2008,Magri,Subramanian2011,Waugh2011,Juniper2011,Juniper2011g,Kulkarni2011,Magri2013d}; energy norms \cite{Nagaraja2009,JosephGeorge2011,George2012}; continuation methods \cite{Jahnke1994,Ananthkrishnan2005,Subramanian2010a,Illingworth2013,Waugh2013b}; weakly nonlinear theory \cite{Wicker1996,Juniper2012a,Subramanian2013a,Ghirardo2013a}; dynamical systems theory, and time-series analysis \cite{Kabiraj2012,Kabiraj2012a,Kabiraj2012b,Kashinath2013}.

These notes give a detailed description of thermo-acoustic sensitivity analysis via adjoint-based approaches, which have been developed recently by Magri and Juniper \cite{Magri2013,Magri2013e,Magri2013c} extending the theory of non-reacting incompressible flows by, among others, Hill \cite{Hill1992}, Giannetti and Luchini \cite{Giannetti2007}, Marquet, Sipp and Jacquin \cite{Marquet2008}.

In this paper we lay out the theoretical foundations of adjoint equations, defining mathematically the adjoint eigenfunction and physically interpreting it as the system's receptivity to open-loop forcing. 
We show how to combine the direct and adjoint eigenfunctions to obtain an exact formula for the first-order eigenvalue change when the system is altered by a generic (small) perturbation. 
The entire theoretical framework is described via two different approaches: 
Continuous Adjoint (CA) and Discrete Adjoint (DA). The former operates on the continuous system, for example, of partial differential equations; whereas the latter operates on the numerically discretized system.  
To show pedagogically how the technique works, we analytically study a simple damped oscillator. 
Then, we apply these theoretical concepts to a 
Rijke tube containing an electrically heated hot wire (gauze)  \cite{Matveev2003a,Balasubramanian2008a,Juniper2011}, also  
considering the mean-flow temperature discontinuity (jump) at the flame's location. 
We show that the optimal stabilizing mechanism is a drag-exerting device placed at the downstream end of the duct, also when the mean-flow temperature jump is modelled. This conclusion corroborates the analysis carried out with no mean-flow temperature jump \cite{Magri2013,Magri2013e}. 
Moreover, we briefly discuss two feedback mechanisms, referring to more detailed studies for in-depth-analysis \cite{Magri2013,Magri2013e}: 
a second hot wire 
and a local smooth variation of the tube cross-sectional area. 
These feedback mechanisms turn out to be effective at changing the frequency of the oscillations, but not the growth rate.
We discuss the role of the mean-flow temperature jump by comparing the solutions with/without temperature jump. 
In the concluding remarks and discussion, we describe the use of these techniques in ongoing research in reduced-order thermo-acoustics.
\section{Definition of the adjoint function}\label{subset:functional_an}
\begin{figure}[h]
\centering \includegraphics[width=14cm]{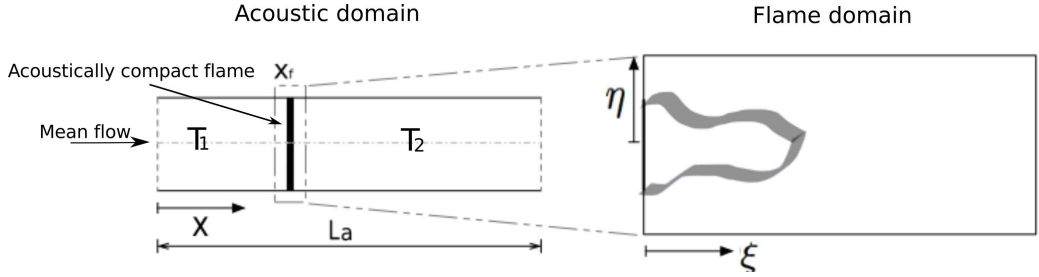}
\caption{Reduced-order thermo-acoustic model with acoustically compact flame and mean-flow temperature jump.}
\label{scheme}
\end{figure}

\textcolor{\ccolor}{We consider reduced-order thermo-acoustic models in which the flame is acoustically treated as a compact monopole source of sound. In these models there are two computational space domains: (i) the one-dimensional domain in which the acoustics are solved and the heat released by the flame is regarded as a pointwise source; (ii) the flame domain, in which the flame is solved and the heat released by chemical reaction is spatially integrated in order to feed back into the acoustic energy equation. 
This creates a feedback loop between the acoustics and the flame. The steady contribution of the heat released by the flame induces a discontinuous change of the mean-flow properties across the flame's location. A schematic of the reduced-order thermo-acoustic model considered is given in figure \ref{scheme}.}

The direct\footnote{The {\it direct} equations are the {\it governing} equations.} and adjoint problems are expressed, respectively, as:
\begin{align} \label{form_pr_da}
&\mathrm{A}\frac{\partial \mathbf{q}}{\partial t}-\mathrm{L}\mathbf{q}=\hat{\mathbf{s}}\exp\left(\sigma_s t\right),\\  \label{form_pr_da2}
&\mathrm{A}^+\frac{\partial^+ \mathbf{q}^+}{\partial t}-\mathrm{L}^+\mathbf{q}^+=0,
\end{align}
where $\hat{\mathbf{s}}$ is the spatially varying part of the forcing term, which is set to zero in this section. 
If eqns. \eqref{form_pr_da},\eqref{form_pr_da2} represent the continuous equations, then $\textrm{A}$, $\textrm{L}$, $\textrm{A}^+$, $\textrm{L}^+$ are operators and $\mathbf{q}=(F,u,p)^T$, where 
$u$ is the acoustic velocity,
$p$ is the acoustic pressure,
and $F$ contains the flame variables\textcolor{\ccolor}{, such as the mixture fraction for diffusion flames}. In this case, the adjoint operators  and equations are analytically derived and then numerically discretized (CA, discretization of Continuous Adjoints). 
If eqns. \eqref{form_pr_da},\eqref{form_pr_da2} represent the numerically discretized systems, then $\textrm{A}$, $\textrm{L}$, $\textrm{A}^+$, $\textrm{L}^+$ are matrices (in bold from now on) and $\mathbf{q}=(\mathbf{G},\boldsymbol{\eta},\boldsymbol{\alpha})^T$.
In this case, the adjoint matrices and functions are directly derived from the numerically discretized direct system (DA, Discrete Adjoints). 

When we follow the CA approach, the adjoint systems are defined through a bilinear form\footnote{It is apparent that the calculation of the adjoint function depends on the choice of the bilinear form.} $[\cdot,\cdot]$, such that:
\begin{align}\label{eq:bf_00}
\left[\mathbf{q}^+,\left(\textrm{A}\frac{\partial}{\partial t}-\textrm{L}\right)\mathbf{q}\right] - \left[\left(\textrm{A}^+\frac{\partial^+}{\partial t}-\textrm{L}^+\right)\mathbf{q}^+,\mathbf{q}\right]=\;\; \textrm{constant},
\end{align}
which, in this paper, defines an inner product\footnote{A discussion of possible definitions of the adjoint operator is reported in the supplementary material of Luchini and Bottaro \cite{Luchini2014}.}.
\textcolor{\ccolor}{For brevity, in this paper we define the following bracket operators that represent inner products:
\begin{align} \label{eq:bf_001}
&\lf\langle\mathbf{a},\mathbf{b}\rr\rangle=\frac{1}{V}\int_V\mathbf{a}^*\cdot\mathbf{b}\;\mathrm{d}V,\\ \label{eq:bf_002}
& \lf\langle\lf\langle\mathbf{a},\mathbf{b}\rr\rangle\rr\rangle=\frac{1}{\partial V}\int_{\partial V}\mathbf{a}^*\cdot\mathbf{b}\;\mathrm{d}\partial V,\\ \label{eq:bf_002}
&\lf[\mathbf{a},\mathbf{b}\rr]=\frac{1}{T} \frac{1}{V}\int_0^T\int_{V}\mathbf{a}^*\cdot\mathbf{b}\;\mathrm{d}V\mathrm{d}t,
\end{align}
where $\mathbf{a}$,$\mathbf{b}$ are arbitrary functions in the function space in which the problem is defined; $V$ is the space domain and $\partial V$ is its boundary; $t$ is the time; \textcolor{\ccolor}{and ${}^*$ is the complex conjugate.}
Therefore, in this paper, the adjoint operator is defined through the following relation:}
\begin{align}\label{eq:bf_01}
\int_T\int_V \mathbf{q}^{+*}\mathbf{\cdot}\left(\mathrm{A}\frac{\partial}{\partial t} - \textrm{L}\right)\mathbf{q}\;\textrm{d}V\mathrm{d}t\;-\;\int_T\int_V \left(\mathrm{A}^+\frac{\partial^+}{\partial t} - \textrm{L}^+\right)^*\mathbf{q}^{+*}\mathbf{\cdot}\mathbf{q}\;\textrm{d}V\textrm{d}t \;\;=\;\; \textrm{constant}.
\end{align}
To find the adjoint operator with the CA approach we have to perform integration by parts of \eqref{eq:bf_01}.
The above relation is an elaboration of the generalized Green's identity \cite{Dennery1996,Magri2013}.
The adjoint boundary / initial conditions, arising from integration by parts of \eqref{eq:bf_01}, are defined such that the constant on the RHS is zero. 
By integration by parts, we find the important result that $-\mathrm{A}\partial/\partial t = \mathrm{A}^+\partial^+/\partial t$. Setting $\mathrm{A}^+=\mathrm{A}^*$, then $-\partial/\partial t = \partial^+/\partial t$. In other words, the adjoint operator must evolve backwards in time for a problem to be well-posed. 

When we follow the DA approach, the adjoint matrix, $L^+_{ij}, $ can be defined through the Euclidean product  (in Einstein's notation)
\begin{align}\label{eq:bf_10}
q^{+*}_i L_{ij} q_j - q_i L^{+*}_{ij} q^{+*}_j=0.
\end{align}
The above terms are scalars, so the transposition does not change the equation.  
Therefore, we take the transpose of the second term and equate it to the first term:
\begin{align}\label{eq:bf_11}
q^{+*}_i L_{ij} q_j=(q_i L^{+*}_{ij} q^{+*}_j)^T=&q^{+*}_i L^{+*}_{ji} q_j,\\ \label{eq:bf_12}
&\implies L^+_{ij}=L_{ji}^*.
\end{align}
This shows that the adjoint matrix is the conjugate-transpose of the direct matrix.
From now on, when we use the DA method, we denote the direct state vector as $\boldsymbol{\chi}$ and the corresponding adjoint vector as $\boldsymbol{\xi}$. 

In Magri and Juniper \cite{Magri2013}  a comparison between the numerical truncation errors between the CA and DA methods is illustrated. \textcolor{\ccolor}{Although the two formulations should converge in principle, it has been shown that convergence is not guaranteed a priori \cite{Vogel1995,Bewley2001,Pierce2004}.}
For the thermo-acoustic system considered in this paper, the DA method is more accurate and easier to implement. 
We show, however, the results obtained via the CA method in order to describe how the method works.

In stability/receptivity analysis, we consider the eigenproblem of \eqref{form_pr_da},\eqref{form_pr_da2}:
\begin{align} \label{form_adj}
&\sigma\mathrm{A}\hat{\mathbf{q}}-\mathrm{L}\hat{\mathbf{q}}=0,\\ \label{form_adj2}
&\sigma^+\mathrm{A}^+\hat{\mathbf{q}}^+ -\mathrm{L}^+\hat{\mathbf{q}}^+=0.
\end{align}
\textcolor{\ccolor}{where $\hat{\qvec}$, $\hat{\qvec}^+$ are the eigenfunctions, and $\sigma$, $\sigma^+$ are the eigenvalues.}
A very important property of the adjoint and direct eigenpairs $\{\sigma_i, \hat{\mathbf{q}}_i\}$, $\{\sigma^+_j, \hat{\mathbf{q}}^+_j\}$ is the bi-orthogonality condition: %
\begin{equation}\label{biocon}
\left(\sigma_i-\sigma^{+*}_j\right) \left\langle \hat{\mathbf{q}}^+_j, \mathrm{A}\hat{\mathbf{q}}_i \right\rangle=0,
\end{equation}
 which states that the inner product $ \left\langle \hat{\mathbf{q}}^+_j, \mathrm{A}\hat{\mathbf{q}_i} \right\rangle$ is zero for every pair of eigenfunctions except when $i=j$, as long as $\sigma_j^+=\sigma^*_j$, in accordance with Salwen and Grosch \cite{Salwen1981}. 
 \textcolor{\ccolor}{This means that the adjoint operator's spectrum is the complex conjugate of the direct operator's spectrum. This information serves as good check when validating adjoint algorithms.}
 \subsection{Physical meaning of the adjoint eigenfunction} \label{rec_th}
%
In this section, we show that the adjoint eigenfunction quantifies the system's receptivity to open-loop forcing. 
Then we give an interpretation of the adjoint eigenfunctions for reduced-order thermo-acoustic systems. 
The receptivity of boundary layers has been calculated from the Orr-Sommerfeld equation by Salwen and Grosch \cite{Salwen1981} and Hill \cite{Hill1995}. 
Another elegant formulation of the receptivity problem, based on the inverse Laplace transform and residues theorem, is described by Giannetti and Luchini \cite[pp. 172--174]{Giannetti2007}. 
A more general approach to the receptivity problem via adjoint equations can be found, among others, in Marino and Luchini \cite[p. 42]{Marino2008}; Meliga \textit{et al.} \cite[p. 605]{Meliga2009}; Sipp \textit{et al.} \cite[p. 10]{Sipp2010}; and Luchini and Bottaro \cite{Luchini2014}.
These studies all concern flow stability.
In these notes, we extend these methods to thermo-acoustic instability
using the formulation by Chandler \cite[pp. 63--68]{Chandler2010}, 
which is sufficiently general for our purposes.
\par
Let $\mathbf{q}$ be a time-dependent state vector defined in a suitable function space and $\textrm{L}$ a linear operator encapsulating the boundary conditions. 
We consider the continuous inhomogeneous linear problem \eqref{form_pr_da}, 
with harmonic forcing at complex frequency, $\sigma_s$, and initial condition $\mathbf{q}(t=0) = \mathbf{q}_0$.
The general solution of this problem is (CA approach):
\begin{align}\label{sol1}
\mathbf{q} = \hat{\mathbf{q}}_s\exp\left(\sigma_st\right) + \mathbf{q}_d + \mathbf{q}_{cs},
\end{align}
where $\hat{\mathbf{q}}_s$ is the spatially varying part of the particular solution,
$\mathbf{q}_d= \sum_j^N\beta_j\hat{\mathbf{q}}_j\exp\left(\sigma_jt\right)$ is the discrete-eigenmodes solution,
and $\mathbf{q}_{cs}$ is the continuous-spectrum solution. 
Oden \cite{Oden1979} and Kato \cite{Kato1980} contain rigorous mathematical treatises of spectral decomposition of linear operators.
Note that an open-loop forcing term, such as $\hat{\mathbf{s}}\exp(\sigma_st)$, does not change the spectrum of the operator.
Assuming that the discrete-eigenmodes and continuous-spectrum 
form a complete basis, 
then the particular and homogenous solutions can be projected onto these spaces. 
Invoking the adjoint eigenfunction and taking advantage of the bi-orthogonality condition \eqref{biocon},
we rearrange \eqref{sol1} as:
\begin{align}\label{solin}
\mathbf{q} = 
&\sum_{j=1}^N\left\langle \hat{\mathbf{q}}^+_j, \mathbf{q}_0\exp\left(\sigma_j t\right) + \hat{\mathbf{s}}\frac{\exp\left(\sigma_s t\right)-\exp\left(\sigma_j t\right)}{\sigma_s-\sigma_j}\right\rangle\frac{\hat{\mathbf{q}}_j}{\left\langle\hat{\mathbf{q}}^+_j,\mathrm{A}\hat{\mathbf{q}}_j\right\rangle} +
\textrm{proj}[\hat{\mathbf{q}}_s,\hat{\mathbf{q}}_{cs}]\exp\left(\sigma_s t\right)+\mathbf{q}_{cs},
\end{align}
where $\textrm{proj}[\hat{\mathbf{q}}_s,\hat{\mathbf{q}}_{cs}]$ is the projection of the forcing term onto the continuous spectrum. 
The solution (\ref{solin}) is valid for a continuous operator 
(e.g. Orr-Sommerfeld) 
in an unbounded or semi-unbounded domain. 
We consider reduced-order thermo-acoustic systems 
in which acoustic and combustion domains are bounded. 
In this case, 
we can assume that there is no continuous spectrum, therefore $\hat{\mathbf{q}}_{cs}=0$. 

The first term of (\ref{solin}) provides a physical interpretation of the adjoint eigenfunction. 
The response of 
the $j^{th}$ component of $\mathbf{q}$ in the long-time limit 
increases 
(i) as
the forcing frequency, $\sigma_s$,
approaches 
the $j^{th}$ eigenvalue, $\sigma_j$,
and (ii)
as the spatial structure of the forcing, $\hat{\mathbf{s}}$, 
approaches
the spatial structure of the adjoint eigenfunction, $\hat{\mathbf{q}}^+_j$. 
For constant amplitude forcing (Re($\sigma_s)=0$)
of a system with one unstable eigenfunction (\textcolor{\ccolor}{Re($\sigma_1$)$>0$})
the linear response (\ref{solin}), in the limit $t\rightarrow\infty$, reduces to
\begin{align}\label{solin2}
\mathbf{q} = 
&\left\langle \hat{\mathbf{q}}^+_1, \mathbf{q}_0-\frac{ \hat{\mathbf{s}}}{\sigma_s-\sigma_1}\right\rangle\frac{\hat{\mathbf{q}}_1}{\left\langle\hat{\mathbf{q}}^+_1,\mathrm{A}\hat{\mathbf{q}}_1\right\rangle}\exp\left(\sigma_1 t\right).
\end{align}
This shows that 
the linear response has 
the frequency/growth rate, $\sigma_1$,
and 
the spatial structure, $\hat{\mathbf{q}}_1$,
of the most unstable direct eigenfunction.
Furthermore, 
the magnitude of this response 
is determined by the extent to which
the spatial structure of the initial conditions, 
$\mathbf{q}_0$,
and the spatial structure of the forcing, 
$\hat{\mathbf{s}}$,
project onto the spatial structure of the adjoint eigenfunction, 
$\hat{\mathbf{q}}^+_1$.
In other words,
the flow behaves as an oscillator 
with an intrinsic frequency, growth rate, and shape
\cite{H&M90}
and the corresponding adjoint shape 
quantifies
the sensitivity of this oscillation
to changes in the spatial structure of the forcing or initial condition.
For constant amplitude forcing acting on a stable system,  
the linear response in the limit $t\rightarrow\infty$ reduces to
\begin{align}\label{solin3}
\mathbf{q} = 
&\sum_{j=1}^N\left\langle \hat{\mathbf{q}}^+_j, \frac{\hat{\mathbf{s}}}{\sigma_s-\sigma_j}\right\rangle\frac{\hat{\mathbf{q}}_j}{\left\langle\hat{\mathbf{q}}^+_j,\mathrm{A}\hat{\mathbf{q}}_j\right\rangle}\exp\left(\sigma_s t\right).
\end{align}
This shows that
the linear response is at the forcing frequency, $\sigma_s$,
and that
the spatial structure 
contains contributions from all eigenfunctions, $\hat{\mathbf{q}}_j$.
Furthermore,
the amplitude of each eigenfunction's contribution
increases 
(i) as $\sigma_s$ approaches one eigenvalue, $\sigma_j$
and
(ii) as the spatial structure of the forcing, $\hat{\mathbf{s}}$,
approaches the spatial structure of that adjoint eigenfunction, $\hat{\mathbf{q}}^+_1$.
This shows that the sensitivity of the response of each mode
to changes in the spatial structure of the forcing (\textit{receptivity})
is quantified by each (corresponding) adjoint eigenfunction, $\hat{\mathbf{q}}^+_j$. 

If we consider the discretized system in the inhomogeneous form (DA approach), seeking solutions of the form $\hat{\boldsymbol{\chi}}\exp({\sigma_st})$ then we obtain:
\begin{equation}\label{rec_disc_1}
(\mathbf{A}\sigma_s-\mathbf{L})\hat{\boldsymbol{\chi}} = \hat{\mathbf{g}} + \mathbf{A}\hat{\boldsymbol{\chi}}(0),
\end{equation}
where $\hat{\mathbf{g}}$ is the discretized source term $\hat{\mathbf{s}}$. In the discretized system the spectrum consists of a finite set of points. Assuming that the eigenvectors form a complete set, the solution is decomposed as follows:
\begin{equation}\label{rec_disc_2}
\boldsymbol{\chi} = \sum_{j=1}^{N}\alpha_j\hat{\boldsymbol{\chi}}_j.
\end{equation}
Substituting eq.~(\ref{rec_disc_2})  back into the discretized version of eq.~(\ref{form_adj}) and premultiplying by the conjugate adjoint eigenvector, $\hat{\boldsymbol{\xi}}^*_j$, (which is the conjugate left eigenvector), we obtain
\begin{equation}\label{rec_disc_3}
\sum_{j=1}^{N}\hat{\boldsymbol{\xi}}^*_j\mathbf{\cdot}(-\mathbf{L}+\mathbf{A}\sigma_s)\alpha_j\hat{\boldsymbol{\chi}}_j = \hat{\boldsymbol{\xi}}^*_j\mathbf{\cdot}(\hat{\boldsymbol{g}}+\mathbf{A}\hat{\boldsymbol{\chi}}(0)).
\end{equation}
Finally, recalling the bi-orthogonality condition for generalized eigenvalues problems \eqref{biocon}, we obtain
\begin{equation}\label{rec_disc_4}
\boldsymbol{\chi} = \sum_{j=1}^{N}\frac{1}{\hat{\boldsymbol{\xi}}_j^*\cdot\mathbf{A}\hat{\boldsymbol{\chi}}_j}\frac{1}{\sigma_s-\sigma_j}\hat{\boldsymbol{\xi}}^*_j\mathbf{\cdot}(\hat{\boldsymbol{g}}+\mathbf{A}\hat{\boldsymbol{\chi}}(0))\hat{\boldsymbol{\chi}}_j
\end{equation}
This result is analogous to \eqref{solin} for discretized systems. \\

\textcolor{\ccolor}{
From a constrained optimization point of view, we define a Lagrangian functional as 
\begin{align}\label{eq:4}
\mathcal{L}\lf(\qvec,\qvec_0,\qavec,\qavec_0\rr) = 
\mathcal{J}\lf(\qvec_0,\qvec_{\partial V},\qvec\rr)
- &\lf[\qavec,\mathrm{A}\partial/\partial t\qvec-\mathrm{L}\qvec\rr]
- \lf\langle\qavec_0, \qvec(0) -\qvec_0\rr\rangle+\ldots\nonumber\\
\ldots-\lf\langle\lf\langle\qavec_{\partial V}, \qvec(\partial V) -\qvec_{\partial V}\rr\rangle\rr\rangle,
\end{align}
where $\mathcal{J}$ is the cost functional to optimize and $\qvec_{\partial V}$ is the boundary condition. 
When the cost functional is the eigenvalue, $\mathcal{J}=\sigma$, as in this paper, the eigenproblem is to be constrained. 
The first variation of $\mathcal{L}$,  along the generic direction $\dtqvec$ is defined through the G$\hat{\mathrm{a}}$teaux derivative, as
\begin{align}\label{eq:6}
\frac{\delta\mathcal{L}}{\dqvec}\dtqvec = \lim_{\ep\rightarrow0} \frac{\mathcal{L}\lf(\qvec+\ep\dtqvec\rr) - \mathcal{L}\lf(\qvec\rr)}{\ep}.
\end{align}
By imposing the first variations of $\mathcal{L}$ with respect to the state vector, $\qvec$, to be zero, we define the adjoint equations \eqref{form_adj2} whose
}
eigenfunctions can be regarded as Lagrange multipliers from a constrained optimization perspective \cite{Gunzbur}. 
Therefore, $u^+$ is the Lagrange multiplier of the acoustic momentum equation,  revealing the locations where the thermo-acoustic system is most receptive to 
forcing (e.g. acoustic forcing); $p^+$ is the Lagrange multiplier of the energy equation revealing the locations where the system is most receptive to heat injection; $F^+$ is the Lagrange multiplier of the flame equation. If the flame is a fast-chemistry diffusion flame, then $F^+$ reveals in which regions the flame is most receptive to species injection \cite{Magri2013c}. \textcolor{\ccolor}{The adjoint boundary conditions can be interpreted likewise.}
\section{The role of the adjoint eigenfunction in perturbation theory}\label{sec:adjfun_drift}
 \par
We study the change of the thermo-acoustic stability as a consequence of a generic perturbation to the problem. The aim is to find an analytical  formula for the eigenvalue drift caused by a perturbation to the operator. 
%
%

With the CA approach (Continuous Adjoint) we study the continuous system (before numerical discretization).
The direct operator is perturbed as  $\opL + \epsilon\delta \opL$, and the eigenvalues and eigenfunctions become $\sigma_j + \epsilon\delta \sigma_j$, $\matqh_j + \epsilon\delta \matqh_j$,
and $\matqh^+_j + \epsilon\delta \matqh^+_j$.

We substitute these into the continuous eigenproblem \eqref{form_adj}
and retain terms up to the first order:
\begin{eqnarray}\label{equ_020_3550}
&\sigma_j
\langle
\matqh^+_j 
, 
\mathrm{A}\matqh_j
\rangle
+
 \epsilon\delta \sigma_j
\langle
\matqh^+_j 
, 
\mathrm{A}\matqh_j
\rangle
+
\sigma_j
\langle
\epsilon\delta \matqh^+_j 
, 
\mathrm{A}\matqh_j
\rangle
+
\sigma_j
\langle
\matqh^+_j 
, 
 \mathrm{A}\epsilon\delta \matqh_j
\rangle=\ldots\nonumber\\
 &\ldots= 
\langle
\matqh^+_j
 , 
\opL
\matqh_j
\rangle 
+
\langle
 \epsilon\delta \matqh^+_j
 , 
\opL
\matqh_j
\rangle 
+
\langle
\matqh^+_j
 , 
 \epsilon\delta \opL
\matqh_j
\rangle 
+
\langle
\matqh^+_j
 , 
\opL
 \epsilon\delta \matqh_j
\rangle.
\end{eqnarray}
We know that 
$
\sigma_j
\langle
\matqh^+_j 
, 
\mathrm{A}\matqh_j
\rangle
=
\langle
\matqh^+_j
 , 
\opL
\matqh_j
\rangle 
$
and so 
$
\sigma_j
\langle
\delta \matqh^+_j 
, 
\mathrm{A}\matqh_j
\rangle
=
\langle
\delta \matqh^+_j
 , 
\opL
\matqh_j
\rangle 
$.
Interestingly,
$
\sigma_j
\langle
\matqh^+_j 
, 
\mathrm{A}\delta \matqh_j
\rangle
=
\langle
\matqh^+_j
 , 
\opL
\delta \matqh_j
\rangle 
$
because taking the inner products of 
$\delta \matqh_j$ 
and 
$\opL \delta \matqh_j$ 
with $\matqh^+_j$
extracts only the components that are parallel to $\matqh_j$ (see eqn. \eqref{biocon}),
for which $\opL \matqh_j = \sigma_j \mathrm{A}\matqh_j$. 
This means that the eigenvalue drift is, at first order:
\begin{eqnarray}
\label{equ_020_355}
\delta\sigma_j=
\frac
{
\langle
\matqh^+_j
 , 
\delta \mathrm{L}
\matqh_j
\rangle 
}
{
\langle
\matqh^+_j 
, 
\mathrm{A}
\matqh_j
\rangle
}.
\end{eqnarray}
Note that the denominator is always different from zero because the dimension of the adjoint space is equal to the original space's dimension, under not restrictive conditions \cite{Maddox1988}.

With the DA approach (Discrete Adjoint) we study the discretized system, represented by the matrices $\mathrm{\mathbf{A}}$, $\mathrm{\mathbf{L}}$. 
From \eqref{eq:bf_12} we can infer that the adjoint eigenvector is the conjugate left eigenvector of the system, i.e. $\hat{\boldsymbol{\xi}}_j^*\cdot(\sigma_j\mathrm{\mathbf{A}} - \mathrm{\mathbf{L}}) = 0$. 
The bi-orthogonality property ensues directly from definition of right and left eigenvectors 
\begin{eqnarray}
\label{equ_010_210}
\hat{\boldsymbol{\xi}}_j^*\cdot\mathbf{A}\cdot\hat{\boldsymbol{\chi}}_i = \delta_{ij}.
\end{eqnarray}
Now, let us consider a perturbation to the direct operator, as before, such that the discretized version of
 \eqref{equ_020_3550} becomes:
\begin{eqnarray}
\label{equ_010_215}
\big( (\sigma_i + \epsilon \delta \sigma_i) \mathbf{A} -(\mathrm{\mathbf{L}} + \epsilon \delta \mathrm{\mathbf{L}}) \big)(\hat{\boldsymbol{\chi}}_i+ \epsilon \delta \hat{\boldsymbol{\chi}}_i)= 0.
\end{eqnarray}
At order $\epsilon$ this is:
\begin{eqnarray}
\label{equ_010_220}
( \sigma_i \mathbf{A} -\mathrm{\mathbf{L}} ) \epsilon \delta \hat{\boldsymbol{\chi}}_i
+
(\epsilon \delta \sigma_i \mathbf{A}-\epsilon \delta \mathrm{\mathbf{L}}  )\hat{\boldsymbol{\chi}}_i= 0.
\end{eqnarray}
Now we pre-multiply by the $i^{th}$ adjoint eigenvector:
\begin{eqnarray}
\label{equ_010_225}
\hat{\boldsymbol{\xi}}_i^*\cdot (\sigma_i \mathbf{A}-\mathrm{\mathbf{L}}  ) \epsilon \delta \hat{\boldsymbol{\chi}}_i
+
\hat{\boldsymbol{\xi}}_i^*\cdot( \epsilon \delta \sigma_i \mathbf{A}-\epsilon \delta \mathrm{\mathbf{L}} )\hat{\boldsymbol{\chi}}_i= 0.
\end{eqnarray}
The first term is zero because of \eqref{equ_010_210}. 
The second term becomes:
\begin{eqnarray}
\label{equ_010_225}
\delta \sigma_i\hat{\boldsymbol{\xi}}_i^* \cdot\mathbf{A}\cdot \hat{\boldsymbol{\chi}}_i=
\hat{\boldsymbol{\xi}}_i^* \cdot \mathrm{\mathbf{L}} \hat{\boldsymbol{\chi}}_i,
\end{eqnarray}
which can be rearranged as:
\begin{eqnarray} \label{equ_010_225}
\delta \sigma_i= 
\frac{ \hat{\boldsymbol{\xi}}_i^*\cdot \delta \mathrm{\mathbf{L}} \hat{\boldsymbol{\chi}}_i}
{\hat{\boldsymbol{\xi}}_i^*\cdot \mathbf{A} \cdot \hat{\boldsymbol{\chi}}_i}.
\end{eqnarray}
\par 
Both eqns. \eqref{equ_020_355} and \eqref{equ_010_225} show that, once the perturbation operator/matrix is known, we can evaluate {\it exactly} at first order the eigenvalue drift. To this end, we need to solve for two eigenproblems to obtain the dominant direct (right) and adjoint (left) eigenfunctions (eigenvectors). This greatly reduces the number of computations without affecting the accuracy. \textcolor{\ccolor}{Eqns.  \eqref{equ_020_355}, \eqref{equ_010_225} are well-known results from perturbation methods (see, e.g., \cite{Stewart1990,Hinch1991}). To the authors' best knowledge, the first scientists who applied this result to hydrodynamic stability were Hill \cite{Hill1992} and Giannetti and Luchini \cite{Giannetti2007}. Although the adjoint equation depends on the choice of the bilinear form, as explained previously, eqns. \eqref{equ_020_355} and \eqref{equ_010_225} do not. }
When the perturbation operator $\delta\mathrm{L}$ represents a perturbation of the thermo-acoustic parameters, such as $\beta$ in eqn. \eqref{equ_gov_mom_ndim2}, we label it as a {\it base-state perturbation}. 
Otherwise, the perturbation is called a {\it structural perturbation} (e.g. the one caused by an external feedback mechanism, like a second hot wire, see sec. \ref{Results_th}). 
%
%
%
\section{Understanding the use of the adjoint eigenfunction}
\label{sec_005_005}
\subsection{A pedagogical example: the damped oscillator}
\par
Culick \cite{Culick2006} showed that a generic thermo-acoustic system behaves like a coupled system of oscillators. 
It is worth presenting a pedagogical example of an application of adjoint analysis for which analytical solutions are available.
This is a lightly-damped linear oscillator consisting of a mass-spring-damper system,  
whose displacement, $x$, given the initial conditions, obeys the governing equation:
\begin{equation}\label{equ_005_005}
\frac{\mathrm{d}^2x}{\mathrm{d}t^2} + b\frac{\mathrm{d}x}{\mathrm{d}t}+ cx = 0.
\end{equation}
This second order ODE can be written as two first order ODEs 
by introducing the velocity, $y$:
\begin{align}\label{equ_005_010}
&\frac{\mathrm{d}x}{\mathrm{d}t} =  y, \\ \label{equ_005_015}
&\frac{\mathrm{d}y}{\mathrm{d}t} =  -by -cx.
\end{align}
We define the state vector $\matq$ and the operator $\opL$, which in this case is a matrix of constant coefficients, such that 
\eqref{form_pr_da} can be written as:
\begin{equation}\label{equ_005_020}
\frac{\ord \matq}{\ord t} - \opL \matq = 0,
\end{equation}
where
\begin{eqnarray}
\label{equ_005_025}
\matq =
\left[ 
\begin{array}{c}
x \\
y
\end{array}
\right],
\qquad\qquad
\opL \matq=
\left[ 
\begin{array}{cc}
0 & 1\\
-c & -b 
\end{array}
\right]
\left[ 
\begin{array}{c}
x \\
y
\end{array}
\right].
\end{eqnarray}
We define the adjoint operator, $\opL^+$,
through \eqref{eq:bf_00}, which gives
\begin{eqnarray} \label{equ_005_050}
\int_0^T{x^+}^*(y)
+
{y^+}^*(-by-cx)\;\mathrm{d}t
& = &
\int_0^T{x^+}^*(y)(\opL^+ \matq^+)^*_x x
+
(\opL^+ \matq^+)^*_y y\;\mathrm{d}t,
\\  \label{equ_005_056}
\implies \int_0^T{x^+}^*(y)(-c^*{y^+})x^* 
+
({x^+} -b^*{y^+})y^*\;\mathrm{d}t
& = &
\int_0^T{x^+}^*(y)(\opL^+ \matq^+)_x x^*
+
(\opL^+ \matq^+)_y y^*\;\mathrm{d}t.
\end{eqnarray}
Note that here the bilinear form does not involve spatial integration over $V$ because the problem is governed by ODEs.
By inspection:
\begin{eqnarray}
\label{equ_005_060}
\matq^+=
\left[ 
\begin{array}{c}
x^+ \\
y^+
\end{array}
\right], \qquad\qquad
\opL^+ \matq^+=
\left[ 
\begin{array}{cc}
0 & -c^*\\
1 & -b^* 
\end{array}
\right]
\left[ 
\begin{array}{c}
{x^+} \\
{y^+}
\end{array}
\right],
\end{eqnarray}
so the adjoint governing equations are:
\begin{eqnarray}
\label{equ_005_070}
\textcolor{\ccolor}{-}\frac{\mathrm{d}x^+}{\mathrm{d}t} & = & -c^* y^+ ,\\
\label{equ_005_075}
\textcolor{\ccolor}{-}\frac{\mathrm{d}y^+}{\mathrm{d}t} & = & -b^*y^+ + x^+.
\end{eqnarray}
By comparing with \eqref{equ_005_010},\eqref{equ_005_015}
we see that
the two first order equations
do not create a self-adjoint system\footnote{\textcolor{\ccolor}{Self-adjointness occurs when the direct operator is equal to its adjoint.}}.
%
%
We consider the following direct and adjoint eigenpairs, respectively:
\begin{align}
\label{equ_010_070}
&\sigma_1  = \frac{-b + \sqrt{b^2 - 4c}}{2},\qquad\qquad\sigma_1^+  =  \sigma_1^*,\\ \label{equ_010_080}
&\matqh_1 
=
\left[ 
\begin{array}{c}
\xh \\
\yh
\end{array}
\right]_1
 =  
\left[ 
\begin{array}{c}
2 \\
-b + \sqrt{b^2-4c} 
\end{array}\right],\qquad\qquad
\matqh_1^+ 
=
\left[ 
\begin{array}{c}
\xh^+ \\
\yh^+
\end{array}
\right]_1
 =  
\left[ 
\begin{array}{c}
-2c^* \\
-b^* - \sqrt{b^{*2}-4c^*} 
\end{array}
\right].
\end{align}
where the minus sign in front of the square root in the $\hat{y}^+$ component of $\matqh_1^+$ in \eqref{equ_010_080} arises because the system is assumed to be lightly damped and therefore $b^{*2}-4c^*$ is negative. 
%
\subsubsection{Eigenvalue sensitivities}
We perturb the system \eqref{equ_005_010},\eqref{equ_005_015} with a small feedback mechanism that feeds from $x$ into the first governing equation:
\begin{eqnarray} \label{equ_025_005}
\frac{\mathrm{d}x}{\mathrm{d}t} & = & \epsilon x + y, \\ \label{equ_025_010}
\frac{\mathrm{d}y}{\mathrm{d}t} & = & -by -cx.
\end{eqnarray}
Note that we considered $b$,$c$ to be real, so, $b=b^*$,$c=c^*$. 
The perturbed state is now:
\begin{eqnarray}
\label{equ_025_015}
(\opL + \delta \opL) \matq
 = 
\left[ 
\begin{array}{cc}
\epsilon & 1\\
-c & -b 
\end{array}
\right]
\left[ 
\begin{array}{c}
x \\
y
\end{array}
\right],
\end{eqnarray}
or in other words:
\begin{eqnarray}
\label{equ_025_020}
\delta \opL
 = 
\left[ 
\begin{array}{cc}
\epsilon & 0\\
0 & 0 
\end{array}
\right].
\end{eqnarray}
We can work out the change in eigenvalue by using the formula derived with the aid of the adjoint eigenfunction \eqref{equ_020_355}:
\begin{eqnarray}
\label{equ_025_030}
\delta \sigma_1
& = &
\frac{
\langle
\matqh^{+}_1 , \delta \opL \matqh_1
\rangle 
}
{
\langle
\matqh^+_1 , \matqh_1
\rangle
}
\\
\label{equ_025_035}
& = &
\frac{
\xh^{+ *}_1
\epsilon
\xh_1
}
{
\langle
\matqh^+_1 , \matqh_1
\rangle
}
\\
\label{equ_025_050}
& = &
\epsilon
\left(
\frac
{
1
}
{2}
+
\frac
{
b
}
{2\sqrt{b^{2} - 4c}}
\right)
\end{eqnarray}
As a check, we will work out $\delta \sigma_1$
by solving exactly the perturbed eigenproblem.
We will use the notation 
$\sigma'_j \equiv \sigma_j + \delta \sigma_j$ for convenience:
\begin{align}
\label{equ_025_085}
&\mathrm{det} \big[(\opL + \delta \opL) - \sigma'_j \matI \big]  =  0,
\\
\label{equ_025_090}
&\mathrm{det}\left[
\begin{array}{cc}
\epsilon -\sigma'_j & 1\\
-c & -b -\sigma'_j 
\end{array}
\right]
=  0.
\\
\label{equ_025_095}
&(\sigma'_j - \epsilon) (\sigma'_j + b) + c  =  0.
\end{align}
Therefore:
\begin{eqnarray}
\label{equ_025_105}
\sigma_1' & = & 
\frac{-(b-\epsilon) + \sqrt{(b-\epsilon)^2 - 4(c-\epsilon b)}}{2} \\
\label{equ_025_110}
\sigma_2' & = & 
\frac{-(b-\epsilon) - \sqrt{(b-\epsilon)^2 - 4(c-\epsilon b)}}{2}
\end{eqnarray}
To calculate the sensitivity to the perturbation, we differentiate with respect to $\epsilon$
\begin{eqnarray}
\label{equ_025_115}
\frac{\ord}{\ord \epsilon}
\big((b-\epsilon)^2 - 4(c-\epsilon b)\big)^{1/2}
 = 
\frac{
-(b-\epsilon)
+2b
}{\big((b-\epsilon)^2 - 4(c-\epsilon b)\big)^{1/2}}.
\end{eqnarray}
So the Taylor expansion of \eqref{equ_025_105}
around $\epsilon = 0$, at first order, gives:
\begin{eqnarray}
\label{equ_025_125}
\sigma_1'
=
\frac{-b + \sqrt{b^2 - 4c}}{2} 
+ 
\epsilon
\left(
\frac{1}{2}
+
\frac{
-(b)
+2b
}{2\big(b^2 - 4c\big)^{1/2}}
\right),
\end{eqnarray}
and therefore the eigenvalue drift is
\begin{eqnarray}
\label{equ_025_130}
\delta \sigma_1
=
\epsilon
\left(
\frac{1}{2}
+
\frac{b}{2\sqrt{b^2 - 4c}}
\right),
\end{eqnarray}
which is the same as \eqref{equ_025_050}, as we wished to show. 
\subsection{Electrically heated Rijke tube with mean-flow temperature jump}\label{Results_th}
The thermo-acoustic system used to demonstrate the adjoint framework is a Rijke tube heated by 
an electrical hot wire (gauze) \cite{Matveev2003a}. 
A full description of such a system - with no mean-flow temperature jump - with relevant non-dimensionalization is given by 
Balasubramanian and Sujith \cite{Balasubramanian2008a}
and Juniper \cite{Juniper2011}.

One-dimensional acoustic waves occur on top of a \emph{mean flow}, which undergoes a discontinuity of its uniform properties across the heat source (see figure \ref{scheme}).  Only the mean-flow pressure does not undergo a discontinuity in the low Mach number limit \cite{Dowling1995,Nicoud2009}.   
The acoustic momentum, energy equations and heat-release law are, respectively:
\begin{align}
\label{equ_gov_mom_ndim}
&\rho\frac{\partial u}{\partial t} + \frac{\partial p}{\partial x} = 0,\\ \label{equ_gov_mom_ndim2}
&\frac{\partial p}{\partial t} + \frac{\partial u}{\partial x}
+ \zeta p
- \beta\dot{q}\delta_f= -u_c\frac{1}{\gamma}\frac{\partial \gamma}{\partial x}\theta_c,\\
&\dot{q}=
\frac{\sqrt{3}}{2}\left[u_f(t) - \tau\left(\frac{\partial u(t)}{\partial t}\right)_{f}\right],\label{qnl}
\end{align}
where $u$, $p$ are the non-dimensional acoustic velocity and pressure. The heat-transfer coefficient, $\beta$, is assumed to be constant and its complete expression, encapsulating the hot wire's properties and ambient conditions, is reported in \cite{Juniper2011}.
\textcolor{\ccolor}{The acoustic velocity has been non-dimenensionalized with the mean-flow velocity; the acoustic pressure with $\kappa M_1\bar{p}$, where $\kappa$ is the heat capacity ratio, $M_1$ is the cold-flow Mach number and $\bar{p}$ the mean-flow pressure; the abscissa with the duct length, ${L_a}$; the time with ${L}_a/\bar{c}_1$, where $\bar{c}_1$ is the cold-flow speed of sound.}
The heat-release rate, $\dot{q}$, is the linearized version  of the nonlinear time-delayed law proposed by Heckl \cite{Heckl1990}, in which the subscript $f$ means that the variable is evaluated at the hot wire's location $x=x_f$ ($\delta_f\equiv\mathit{\delta(x-x_f)}$ is the Dirac delta distribution). The time delay between the pressure and heat-release oscillations is modelled by the constant coefficient, $\tau$. %
This linearization has been performed both in amplitude and time. Eqn. \eqref{qnl} holds providing that $|u_f(t-\tau)|\ll1$ and $\tau\ll2/K$, 
where $K$ is the number of Galerkin modes considered in the numerical discretization \cite{Juniper2011,Magri2013,Magri2013e}. 
The non-dimensional density is $\rho=\rho_1$ when $x<x_f$ and $\rho=\rho_2$ when $x>x_f$. 
The positive mean-flow temperature jump, induced by the heat transferred to the mean-flow, makes the density ratio $\rho_2/\rho_1<1$ because of the ideal-gas law.
$\gamma\equiv A(x)/A_0$, where $A(x)$ is the area at location $x$ and $A_0$ is a reference area; and $\theta_c$ is 1 at $x=x_c$ and zero elsewhere. If the duct is straight, then $\gamma=0$.
As shown by Magri and Juniper \cite{Magri2013e}, a {\it local smooth cross-sectional area variation}, defined such that $\partial\gamma/\partial x \theta_c$ is finite, can be regarded as a passive feedback mechanism.  
We assume that the area of the duct is constant except at location $x=x_c$,
where there is a small smooth change in the area.
At the ends of the tube, $p$ and $\partial u/ \partial x$ are both set to zero.  
\textcolor{\ccolor}{Dissipation and end losses are modelled by the modal damping  $\zeta=c_1j^2+c_2j^{0.5}$ used by Matveev \cite{Matveev2003}, based on models by Landau and Lifshitz \cite{Landau1987}, where $j$ is the $j^{th}$ acoustic mode. The quadratic term represents the losses at the end of the tube, while the square-rooted term represents the losses in the viscous/thermal boundary layers.}
%

%
\par
\begin{figure}[h]
\centering \includegraphics[width=10cm]{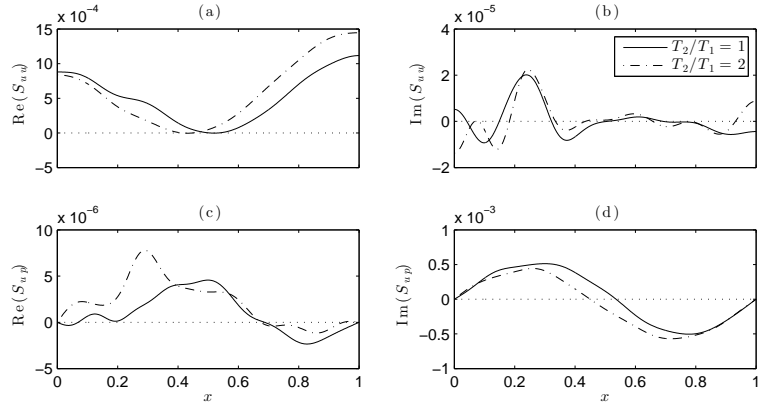}
\caption{Two significant components of the structural sensitivity tensor, which quantify the effect of feedback mechanisms, placed at $x$, on the linear growth rate (left frames) and angular frequency (right frames). Solutions with no mean-flow temperature jump (solid lines) and with temperature jump (dash-dotted line). $c_1=0.01$, $c_2=0.001$, $\tau=0.01$, $\beta=0.433$, $x_f=0.25$, \textcolor{\ccolor}{10 Galerkin modes are used for numerical discretization}.}
\label{fig:RijkeSS}
\end{figure}

\textcolor{\ccolor}{The partial differential equations (\ref{equ_gov_mom_ndim}), (\ref{equ_gov_mom_ndim2}) 
are discretized into a set of ordinary differential equations
by choosing a basis that matches the boundary conditions and the discontinuity condition at the flame. 
The Galerkin method, which is a weak-form method, ensures that in the subspace where the solution is discretized the error is orthogonal to the chosen basis, so that the solution is an optimal weak-form solution.
The pressure, $p$, and velocity, $u$, are expressed by separating the time and space dependence, as follows
\begin{eqnarray}\label{equ_gal_p1}
p(x,t)=\sum_{j = 1}^{K} 
\begin{cases}
\alpha_j^{(1)}(t)\Psi^{(1)}_j(x), & 0\leq x<x_f,\\
\alpha_j^{(2)}(t)\Psi^{(2)}_j(x), &x_f<x\leq1.
\end{cases}
\end{eqnarray}
\begin{eqnarray}\label{equ_gal_u1}
u(x,t)=\sum_{j = 1}^{K} 
\begin{cases}
\eta_j^{(1)}(t)\Phi^{(1)}_j(x), & 0\leq x<x_f,\\
\eta_j^{(2)}(t)\Phi^{(2)}_j(x), &x_f<x\leq1.
\end{cases}
\end{eqnarray}
The system \eqref{equ_gov_mom_ndim}, \eqref{equ_gov_mom_ndim2} reduces to the D'Alembert equation when $\zeta=0$ and $\beta_T=0$
\begin{equation}\label{dalembert}
\frac{\partial^2 p}{\partial t^2} - \frac{1}{\rho}\frac{\partial^2 p}{\partial x^2}=0.
\end{equation}
The following procedure is applied to find the bases for $u$ and $p$:
\begin{enumerate} 
\item substitute the decomposition \eqref{equ_gal_p1} into \eqref{dalembert} to find 
the pressure eigenfunctions $\Psi^{(1)}_j(x)$, $\Psi^{(2)}_j(x)$;
\item  substitute the pressure eigenfunctions  $\Psi^{(1)}_j(x)$, $\Psi^{(2)}_j(x)$ into the momentum equation \eqref{equ_gov_mom_ndim} to find the velocity eigenfunctions  $\Phi^{(1)}_j(x)$, $\Phi^{(2)}_j(x)$;
\item impose the jump condition at the discontinuity, for which $p(x\rightarrow x^-_f)=p(x\rightarrow x^+_f)$ and $u(x\rightarrow x^-_f)=u(x\rightarrow x^+_f)$, to find the relations between $\alpha_j^{(1)}$, $\alpha_j^{(2)}$, $\eta_j^{(1)}$, $\eta_j^{(2)}$. 
\end{enumerate}
 Similarly to Zhao \cite{Zhao2012}, these steps give
 \begin{eqnarray}\label{equ_gal_p11}
p(x,t)=\sum_{j = 1}^{K} 
\begin{cases}
-\alpha_j(t)\sin\left(\omega_j\sqrt{\rho_1}x\right), & 0\leq x<x_f,\\
-\alpha_j(t)\left(\frac{\sin\gamma_j}{\sin\beta_j}\right)\sin\left(\omega_j\sqrt{\rho_2}(1-x)\right), &x_f<x\leq1,
\end{cases}
\end{eqnarray}
\begin{eqnarray}\label{equ_gal_u11}
u(x,t)=\sum_{j = 1}^{K} 
\begin{cases}
\eta_j(t)\frac{1}{\sqrt{\rho_1}}\cos\left(\omega_j\sqrt{\rho_1}x\right), & 0\leq x<x_f,\\
- \eta_j(t)\frac{1}{\sqrt{\rho_2}}\left(\frac{\sin\gamma_j}{\sin\beta_j}\right)\cos\left(\omega_j\sqrt{\rho_2}(1-x)\right), &x_f<x\leq1.
\end{cases}
\end{eqnarray}
where
\begin{equation}\label{cosgcosb}
\gamma_j \equiv \omega_j\sqrt{\rho_1}x_f, \;\;\;\;\beta_j \equiv\omega_j\sqrt{\rho_2}(1-x_f).
\end{equation}
Point 3 of the previous procedure provides the equation for the natural acoustic frequencies $\omega_j$
\begin{equation}\label{eig_val}
\sin\beta_j\cos\gamma_j+\cos\beta_j\sin\gamma_j\sqrt{\frac{\rho_1}{\rho_2}}=0.
\end{equation}
The full description and implementation of this method is available in Magri and Juniper \cite{Magri2013c} based on the numerical model of Zhao \cite{Zhao2012}. }
The continuous adjoint equations of the straight Rijke tube, derived via \eqref{eq:bf_01}, are
\begin{align}
\label{ad1}
&\rho\frac{\partial u^+}{\partial t} + \frac{\partial p^+}{\partial x}  +\frac{\sqrt{3}}{2}\beta\left(p_f^+ +\tau \left(\frac{\partial p^+}{\partial t}\right)_f\right)\delta_f=0,\\ \label{ad2}
&\frac{\partial u^+}{\partial x} + \frac{\partial p^+}{\partial t} - \zeta p^+=0.
\end{align}
Note that \eqref{ad1} differs from the adjoint equations presented in previous work \cite{Magri2013,Magri2013b,Magri2013e} because of the presence of $\rho$, which contains the information about the mean-flow temperature jump. 
\textcolor{\ccolor}{The localized smooth area variation term, $-u_c/\gamma\partial\gamma/\partial x \theta_c$, does not appear in the adjoint equations \eqref{ad1},\eqref{ad2}. This is because this term is viewed as a forcing term of the energy equation \eqref{equ_gov_mom_ndim2} and the adjoint equations are defined with respect to the homogenous direct equations (see \eqref{form_pr_da},\eqref{form_pr_da2},\eqref{eq:bf_00}).}
The direct and conjugate adjoint eigenfunctions are arranged as column vectors $\mathit{[\hat{u}, \hat{p}]^T}$ and $\mathit{[\hat{u}^{+*}, \hat{p}^{+*}]^T}$, respectively.
The structural sensitivity tensor, defined in Magri and Juniper \cite{Magri2013,Magri2013e}, is 
 %
 \begin{equation}
S\equiv\frac{\delta\sigma}{\delta \mathbf{C}}=\frac{[\hat{u}^{+*}, \hat{p}^{+*}]^T\otimes[\hat{u}, \hat{p}]^T}{\int_{0}^{1} \! (\hat{u}\hat{u}^{+*}  + \hat{p}\hat{p}^{+*}) \mathrm{d} x+\beta\tau\hat{u}_f \hat{p}^{+*}_f}, \label{drift_rik} 
\end{equation}
where $\otimes$ denotes the dyadic product and $\delta\mathbf{C}$ is a matrix of arbitrarily small perturbation coefficients. 
Each component of this structural perturbation tensor 
quantifies the effect of a feedback mechanism between a variable
and a governing equation, as explained in \cite{Magri2013,Magri2013e}.
Therefore we can identify the device, and the location,
that is most effective at changing
the frequency or growth rate of the system just by inspection of the components of the structural sensitivity tensor \eqref{drift_rik}.
Here, we discuss the two most significant mechanisms, given by the components 
\begin{align}\label{Suu}
&S_{uu} = \frac{\hat{u}^{+*}\hat{u}}{\int_{0}^{1} \! (\hat{u}\hat{u}^{+*}  + \hat{p}\hat{p}^{+*}) \mathrm{d} x+\beta\tau\hat{u}_f \hat{p}^{+*}_f}, \\ \label{Sup}
&S_{up} = \frac{\hat{p}^{+*}\hat{u}}{\int_{0}^{1} \! (\hat{u}\hat{u}^{+*}  + \hat{p}\hat{p}^{+*}) \mathrm{d} x+\beta\tau\hat{u}_f \hat{p}^{+*}_f}.  
\end{align}
The reader may refer to  \cite{Magri2013,Magri2013e} for a detailed explanation of the remaining components of the structural sensitivity tensor.
The components \eqref{Suu},\eqref{Sup} are shown in fig. \ref{fig:RijkeSS} as a function of $\mathit{x}$, 
which is the location where the passive device (structural perturbation) acts,  both when $\rho_1/\rho_2=T_2/T_1=1$ (solid line) and  $\rho_1/\rho_2=T_2/T_1=2$ (dash-dot line). 

The component $S_{uu}$ is the eigenvalue's sensitivity to a feedback mechanism proportional to the velocity at a given point and affecting the momentum equation. For example, this could be the (linearized) drag force about an obstacle in the flow, as proposed in \cite{Magri2013,Magri2013e}. 
The real part of $S_{uu}$ (fig.~\ref{fig:RijkeSS}a), being the sensitivity of the system's growth rate, is positive for all values of $\mathit{x}$, 
which means that, whatever value of $\mathit{x}$ is chosen,
the growth rate will decrease
if the forcing is in the opposite direction to the velocity, as it is in drag-exerting devices.
This tells us that the drag force of a mesh will always stabilize the thermo-acoustic oscillations but is most effective if placed at  the downstream end of the duct. This effect is even stronger if the temperature jump is considered.
In summary, this type of feedback greatly affects the growth rate
but hardly affects the frequency (fig.~\ref{fig:RijkeSS}b), in agreement with Dowling \cite{Dowling1995}, who performed stability analysis via classical approaches. 

The component $S_{up}$ is the eigenvalue's sensitivity to a feedback mechanism proportional to the velocity at a given point and affecting the energy equation.
This type of feedback hardly affects the growth rate (fig.~\ref{fig:RijkeSS}c) but greatly affects the frequency (fig.~\ref{fig:RijkeSS}d). 
By inspection of the linearized heat law \eqref{qnl}, we notice that a second hot wire with $\tau=0$ causes this type of feedback,
so this analysis shows that it will be relatively ineffective at stabilizing thermo-acoustic oscillations whereas it will be effective at changing the oscillation frequency. A detailed analysis and physical explanation of this finding is reported in \cite{Magri2013,Magri2013e}.

If $\gamma\not=0$, the RHS of eqn.~\eqref{equ_gov_mom_ndim2} shows that a change in the area can be interpreted as a forcing term, proportional to $-u_c$, acting on the energy equation. In other words, a positive local smooth change of the cross-sectional area is a feedback mechanism acting like a second hot wire with negative $\beta$.
Hence, the structural sensitivity is provided by $-S_{up}$.
This means that where a control hot wire has a stabilizing effect, a positive change in area in the same location has a destabilizing effect, and vice versa. 
%

\textcolor{\ccolor}{It is worth mentioning that the structural sensitivity coefficients depicted in figure \ref{fig:RijkeSS} do not depend on the time delay, $\tau$, as long as it remains small compared with the oscillation period. We performed calculations for time delays from 0 to 0.03 and observed negligible differences (results not shown).}

\section{Concluding remarks and discussion}
The aim of this paper is to show how adjoint sensitivity analysis can be applied to thermo-acoustics. 
We describe the physical meaning of the adjoint eigenfunction in terms of the system's receptivity to
open-loop forcing and show how to combine the direct and adjoint eigenfunctions to obtain an analytical formula for the first-order eigenvalue drift. 
We improve the sensitivity analysis of an electrical heated Rijke tube of Magri and Juniper \cite{Magri2013,Magri2013e}
by including the effect of the mean-flow temperature jump in the acoustics. 
Devices exerting a drag force on the fluid
have the biggest effect on the growth rate, whether or not the temperature jump is modelled. 
For the models used in this paper, the optimal place for such drag-exerting device is found to be at the downstream end of the duct. The presence of the  mean-flow temperature jump makes the system even more sensitive to such a stabilizing device. 
In general, including the mean-flow temperature jump alters markedly the shape of the spatial sensitivities to second hot wires and local smooth cross-sectional area variations. \\

The results in this paper are for a simple thermo-acoustic model and are only as accurate as the model itself. The adjoint-based techniques, however, can readily be applied to more realistic models, as long as they can be linearized. This could quickly reveal, for example, 
the best position for an acoustic damper in a complex acoustic network, the optimal change in the flame shape and \textcolor{\ccolor}{it could also suggest strategies for open loop control.}
The usefulness of adjoint techniques applied to thermo-acoustics is that,
in a few calculations, one can predict accurately
how the growth rate and frequency of thermo-acoustic oscillations 
are affected either by all possible passive control elements in the system 
or by all possible changes to its base state. \\

L.M. would like to thank C. Hennekinne for his comments on this paper. 
L.M.'s PhD is supported by the European Research Council through Project ALORS 2590620. 
The Cambridge Philosophical Society (UK) is gratefully acknowledged for having partially covered travel costs for $n^3l$ conference, 2013.

%
%
%
\bibliography{library} 

\begin{thebibliography}{10}

\bibitem{Rayleigh1878}
{Lord Rayleigh}.
\newblock {The explanation of certain acoustical phenomena}.
\newblock {\em Nature}, 18:319--321, 1878.

\bibitem{Lieuwen2005}
T.~C. Lieuwen and V.~Yang.
\newblock {\em {Combustion Instabilities in Gas Turbine Engines: Operational
  Experience, Fundamental Mechanisms, and Modeling}}.
\newblock American Institute of Aeronautics and Astronautics, Inc., 2005.

\bibitem{Culick2006}
F.~E.~C. Culick.
\newblock {\em {Unsteady motions in combustion chambers for propulsion
  systems}}.
\newblock RTO AGARDograph AG-AVT-039, North Atlantic Treaty Organization, 2006.

\bibitem{Lieuwen2012}
T.~C. Lieuwen.
\newblock {\em {Unsteady combustor physics}}.
\newblock Cambridge University Press, 2012.

\bibitem{Balasubramanian2008a}
K.~Balasubramanian and R.~I. Sujith.
\newblock {Thermoacoustic instability in a Rijke tube: Non-normality and
  nonlinearity}.
\newblock {\em Physics of Fluids}, 20(4):044103, 2008.

\bibitem{Balasubramanian2008}
K.~Balasubramanian and R.~I. Sujith.
\newblock {Non-normality and nonlinearity in combustion-acoustic interaction in
  diffusion flames}.
\newblock {\em Journal of Fluid Mechanics}, 594:29--57, December 2008.

\bibitem{Magri}
K.~Balasubramanian and R.~I. Sujith.
\newblock {Non-normality and nonlinearity in combustion-acoustic interaction in
  diffusion flames -- CORRIGENDUM}.
\newblock {\em Journal of Fluid Mechanics}, 733:680--680, 2013.

\bibitem{Subramanian2011}
P.~Subramanian and R.~I. Sujith.
\newblock {Non-normality and internal flame dynamics in premixed flame-acoustic
  interaction}.
\newblock {\em Journal of Fluid Mechanics}, 679(2011):315--342, May 2011.

\bibitem{Waugh2011}
I.~C. Waugh and M.~P. Juniper.
\newblock {Triggering in a thermoacoustic system with stochastic noise}.
\newblock {\em International Journal of Spray and Combustion Dynamics},
  3(3):225--242, September 2011.

\bibitem{Juniper2011}
M.~P. Juniper.
\newblock {Triggering in the horizontal Rijke tube: non-normality, transient
  growth and bypass transition}.
\newblock {\em Journal of Fluid Mechanics}, 667:272--308, November 2011.

\bibitem{Juniper2011g}
M.~P. Juniper.
\newblock {Transient growth and triggering in the horizontal Rijke tube}.
\newblock {\em International journal of spray and combustion dynamics},
  3(3):209--224, 2011.

\bibitem{Kulkarni2011}
R.~Kulkarni, K.~Balasubramanian, and R.~I. Sujith.
\newblock {Non-normality and its consequences in active control of
  thermoacoustic instabilities}.
\newblock {\em Journal of Fluid Mechanics}, 670:130--149, February 2011.

\bibitem{Magri2013d}
L.~Magri, K.~Balasubramanian, R.~I. Sujith, and M.~P. Juniper.
\newblock {Non-normality in combustion-acoustic interaction in diffusion
  flames: a critical revision}.
\newblock {\em Journal of Fluid Mechanics}, 733:681--684, 2013.

\bibitem{Nagaraja2009}
S.~Nagaraja, K.~Kedia, and R.~I. Sujith.
\newblock {Characterizing energy growth during combustion instabilities:
  Singularvalues or eigenvalues?}
\newblock {\em Proceedings of the Combustion Institute}, 32(2):2933--2940,
  2009.

\bibitem{JosephGeorge2011}
K.~{Joseph George} and R.~I. Sujith.
\newblock {On Chu's disturbance energy}.
\newblock {\em Journal of Sound and Vibration}, 330(22):5280--5291, October
  2011.

\bibitem{George2012}
K.~{Joseph George} and R.I. Sujith.
\newblock {Disturbance energy norms: A critical analysis}.
\newblock {\em Journal of Sound and Vibration}, 331(7):1552--1566, March 2012.

\bibitem{Jahnke1994}
C.~C. Jahnke and F.~E.~C. Culick.
\newblock {Application of dynamical systems theory to nonlinear combustion
  instabilities}.
\newblock {\em Journal of Propulsion and Power}, 10(4):508--517, July 1994.

\bibitem{Ananthkrishnan2005}
N.~Ananthkrishnan, S.~Deo, and F.E.C. Culick.
\newblock {Reduced-Order Modeling and Dynamics of Nonlinear Acoustic Waves in a
  Combustion Chamber}.
\newblock {\em Combustion Science and Technology}, 177(2):221--248, January
  2005.

\bibitem{Subramanian2010a}
P.~Subramanian, S.~Mariappan, R.~I. Sujith, and P.~Wahi.
\newblock {Bifurcation analysis of thermoacoustic instability in a horizontal
  Rijke tube}.
\newblock {\em International Journal of Spray and Combustion Dynamics},
  2(4):325--355, 2010.

\bibitem{Illingworth2013}
S.~J. Illingworth, I.~C. Waugh, and M.~P. Juniper.
\newblock {Finding thermoacoustic limit cycles for a ducted Burke-Schumann
  flame}.
\newblock {\em Proceedings of the Combustion Institute}, 34(1):911--920,
  January 2013.

\bibitem{Waugh2013b}
I.~Waugh, S.~Illingworth, and M.~Juniper.
\newblock {Matrix-free continuation of limit cycles for bifurcation analysis of
  large thermoacoustic systems}.
\newblock {\em Journal of Computational Physics}, 240:225--247, May 2013.

\bibitem{Wicker1996}
J.~M. Wicker, W.~D. Greene, S.-III Kim, and V.~Yang.
\newblock {Triggering of longitudinal combustion instabilities in rocker
  motors: nonlinear combustion response}.
\newblock {\em Journal of Propulsion and Power}, 12(6):1148--1158, 1996.

\bibitem{Juniper2012a}
M.~P. Juniper.
\newblock {Triggering in thermoacoustics}.
\newblock {\em International Journal of Spray and Combustion Dynamics},
  4(3):217--238, September 2012.

\bibitem{Subramanian2013a}
P.~Subramanian, R.~I. Sujith, and P.~Wahi.
\newblock {Subcritical bifurcation and bistability in thermoacoustic systems}.
\newblock {\em Journal of Fluid Mechanics}, 715:210--238, January 2013.

\bibitem{Ghirardo2013a}
G.~Ghirardo and M.~P. Juniper.
\newblock {Azimuthal instabilities in annular combustors: standing and spinning
  modes}.
\newblock {\em Proc. R. Soc. Lond. A}, 469(2157):20130232, 2013.

\bibitem{Kabiraj2012}
L.~Kabiraj, A.~Saurabh, P.~Wahi, and R.~I. Sujith.
\newblock {Route to chaos for combustion instability in ducted laminar premixed
  flames}.
\newblock {\em Chaos}, 22(2):023129, 2012.

\bibitem{Kabiraj2012a}
L.~Kabiraj and R.~I. Sujith.
\newblock {Bifurcations of self-excited ducted laminar premixed flames}.
\newblock {\em Journal of Engineering of Gas Turbines and Power}, 134:031502,
  2012.

\bibitem{Kabiraj2012b}
L.~Kabiraj and R.~I. Sujith.
\newblock {Nonlinear self-excited thermoacoustic oscillations: intermittency
  and flame blow-out}.
\newblock {\em Journal of Fluid Mechanics}, 713:376--397, 2012.

\bibitem{Kashinath2013}
K.~Kashinath, I.~C. Waugh, and M.~P. Juniper.
\newblock {Nonlinear self-excited thermoacoustic oscillations of a ducted
  premixed flame: bifurcations and routes to chaos}.
\newblock {\em Journal of Fluid Mechanics, to be submitted}, 2013.

\bibitem{Magri2013}
L.~Magri and M.~P. Juniper.
\newblock {Sensitivity analysis of a time-delayed thermo-acoustic system via an
  adjoint-based approach}.
\newblock {\em Journal of Fluid Mechanics}, 719:183--202, 2013.

\bibitem{Magri2013e}
L.~Magri and M.~P. Juniper.
\newblock {A Theoretical Approach for Passive Control of Thermoacoustic
  Oscillations: Application to Ducted Flames}.
\newblock {\em Journal of Engineering for Gas Turbines and Power},
  135(9):091604, August 2013.

\bibitem{Magri2013c}
L.~Magri and M.~P. Juniper.
\newblock {Global mode, receptivity and sensitivity analysis of diffusion
  flames coupled with acoustics}.
\newblock {\em J. Fluid Mech. submitted}, 2013.

\bibitem{Hill1992}
D.~C. Hill.
\newblock {A theoretical approach for analyzing the restabilization of wakes}.
\newblock {\em NASA Technical memorandum 103858}, 1992.

\bibitem{Giannetti2007}
F.~Giannetti and P.~Luchini.
\newblock {Structural sensitivity of the first instability of the cylinder
  wake}.
\newblock {\em Journal of Fluid Mechanics}, 581:167--197, 2007.

\bibitem{Marquet2008}
O.~Marquet, D.~Sipp, and L.~Jacquin.
\newblock {Sensitivity analysis and passive control of cylinder flow}.
\newblock {\em Journal of Fluid Mechanics}, 615:221--252, November 2008.

\bibitem{Matveev2003a}
K.~Matveev.
\newblock {\em {Thermoacoustic Instabilities in the Rijke Tube: Experiments and
  Modeling}}.
\newblock PhD thesis, Caltech Institute of Technology, 2003.

\bibitem{Luchini2014}
P.~Luchini and A.~Bottaro.
\newblock {Adjoint equations in stability analysis}.
\newblock {\em Ann. Rev. Fluid Mech.}, 46:1--30, 2014.

\bibitem{Dennery1996}
P.~Dennery and A.~Krzywicky.
\newblock {\em {Mathematics for Physicists}}.
\newblock Dover Publications, Inc., 1996.

\bibitem{Vogel1995}
C.~R. Vogel and J.~G. Wade.
\newblock {Analysis of Costate Discretizations in Parameter Estimation for
  Linear Evolution Equations}.
\newblock {\em SIAM Journal on Control and Optimization}, 33(1):227--254, 1995.

\bibitem{Bewley2001}
T.~Bewley.
\newblock {Flow control: new challanges for a new Renaissance}.
\newblock {\em Prog. Aerospace Sci.}, 37:21--58, 2001.

\bibitem{Pierce2004}
N.~A. Pierce and M.~B. Giles.
\newblock {Adjoint and defect error bounding and correction for functional
  estimates}.
\newblock {\em Journal of Computational Physics}, 200(2):769--794, November
  2004.

\bibitem{Salwen1981}
H.~Salwen and C.~E. Grosch.
\newblock {The continuous spectrum of the Orr-Sommerfeld equation. Part 2.
  Eigenfunction expansions}.
\newblock {\em Journal of Fluid Mechanics}, 104:445--465, April 1981.

\bibitem{Hill1995}
D.~C. Hill.
\newblock {Adjoint systems and their role in the receptivity problem for
  boundary layers}.
\newblock {\em Journal of Fluid Mechanics}, 292:183--204, April 1995.

\bibitem{Marino2008}
L.~Marino and P.~Luchini.
\newblock {Adjoint analysis of the flow over a forward-facing step}.
\newblock {\em Theoretical and Computational Fluid Dynamics}, 23(1):37--54,
  November 2009.

\bibitem{Meliga2009}
P.~Meliga, J.-M. Chomaz, and D.~Sipp.
\newblock {Unsteadiness in the wake of disks and spheres: Instability,
  receptivity and control using direct and adjoint global stability analyses}.
\newblock {\em Journal of Fluids and Structures}, 25(4):601--616, May 2009.

\bibitem{Sipp2010}
D.~Sipp, O.~Marquet, P.~Meliga, and A.~Barbagallo.
\newblock {Dynamics and Control of Global Instabilities in Open-Flows: A
  Linearized Approach}.
\newblock {\em Applied Mechanics Reviews}, 63(3):030801, 2010.

\bibitem{Chandler2010}
G.~J. Chandler.
\newblock {\em {Sensitivity analysis of low-density jets and flames}}.
\newblock PhD thesis, University of Cambridge, 2010.

\bibitem{Oden1979}
J.~T. Oden.
\newblock {\em {Applied functional analysis}}.
\newblock Prentice-Hall, Inc, 1979.

\bibitem{Kato1980}
T.~Kato.
\newblock {\em {Perturbation theory for linear operators}}.
\newblock Springer Berlin / Heidelberg, New York, 2nd edition, 1980.

\bibitem{H&M90}
P~Huerre and P~Monkewitz.
\newblock {Local and global instabilities of spatially developing flows}.
\newblock {\em Ann. Rev. Fluid Mech.}, 22:473--537, 1990.

\bibitem{Gunzbur}
M.~D. Gunzburger.
\newblock {\em {Inverse design and optimisation methods. Introduction into
  mathematical aspects of flow control and optimization}}.
\newblock von Karman Insitute for Fluid Dynamics, Lecture Series 1997-05.,
  1997.

\bibitem{Maddox1988}
I.~J. Maddox.
\newblock {\em {Elements of Functional Analysis}}.
\newblock Cambridge University Press, 2nd edition, 1988.

\bibitem{Stewart1990}
G.~W. Stewart and J.-G. Sun.
\newblock {\em {Matrix Perturbation Theory}}.
\newblock Academic press, Inc, 1990.

\bibitem{Hinch1991}
E.~J. Hinch.
\newblock {\em {Perturbation Methods}}.
\newblock Cambridge University Press, 1991.

\bibitem{Dowling1995}
A.~P. Dowling.
\newblock {The calculation of thermoacoustic oscillations}.
\newblock {\em Journal of sound and vibration}, 180(4):557--581, 1995.

\bibitem{Nicoud2009}
F.~Nicoud and K.~Wieczorek.
\newblock {About the zero Mach number assumption in the calculation of
  thermoacoustic instabilities}.
\newblock {\em International Journal of Spray and Combustion Dynamics},
  1(1):67--111, March 2009.

\bibitem{Heckl1990}
M.~A. Heckl.
\newblock {Non-linear acoustic effects in the Rijke tube}.
\newblock {\em Acustica}, 72:63--71, 1990.

\bibitem{Matveev2003}
K.~I. Matveev and F.~E.~C. Culick.
\newblock {A model for combustion instability involving vortex shedding}.
\newblock {\em Combustion Science and Technology}, 175:1059--1083, 2003.

\bibitem{Landau1987}
L.~D. Landau and E.~M. Lifshitz.
\newblock {\em {Fluid Mechanics}}.
\newblock Pergamon Press, second edition, 1987.

\bibitem{Zhao2012}
D.~Zhao.
\newblock {Transient growth of flow disturbances in triggering a Rijke tube
  combustion instability}.
\newblock {\em Combustion and Flame}, 159(6):2126--2137, June 2012.

\bibitem{Magri2013b}
L.~Magri and M.~P. Juniper.
\newblock {A novel theoretical approach to passive control of thermo-acoustic
  oscillations: application to ducted heat sources}.
\newblock In {\em Proceedings of ASME Turbo Expo GT2013-94344}, 2013.

\end{thebibliography}
\end{document}